\documentclass[12pt,preprint]{aastex}

\usepackage{graphicx}
\usepackage{rotating}

\shorttitle{}
\shortauthors{Nesvorn\'y et al.}

\begin{document}
\baselineskip 19.pt
\title{Isotopic Trichotomy of Main Belt Asteroids from Implantation of Outer Solar System Planetesimals}
\author{David Nesvorn\'y$^{1,*}$, Nicolas Dauphas$^2$, David Vokrouhlick\'y$^3$, Rogerio
Deienno$^1$, Timo Hopp$^4$} 
\affil{(1) Department of Space Studies, 
Southwest Research Institute, 1050 Walnut St., Suite 300,  Boulder, CO 80302, United States}
\affil{(2) Origins Laboratory, Department of the Geophysical Sciences, The University of Chicago, 
5734 S. Ellis Avenue
Chicago, Illinois 60637, United States}
\affil{(3) Institute of Astronomy, Charles University, V Hole\v{s}ovi\v{c}k\'ach 2, CZ–18000 Prague
  8, Czech Republic}
\affil{(4) Max Planck Institute for Solar System Research, Planetary Science Department, 
Justus-von-Liebig-Weg 3, 37077 G\"{o}ttingen, Germany}
\affil{*e-mail:davidn@boulder.swri.edu}

\begin{abstract}
Recent analyses of samples from asteroid (162173) Ryugu returned by JAXA's Hayabusa2 mission 
suggest that Ryugu and CI chondrites formed in the same region of the protoplanetary disk, in
a reservoir that was isolated from the source regions of other carbonaceous (C-type) asteroids.
Here we conduct $N$-body simulations in which CI planetesimals are assumed to have formed in 
the Uranus/Neptune zone at $\sim15$--25 au from the Sun. We show that CI planetesimals are
scattered by giant planets toward the asteroid belt where their orbits can be circularized by 
aerodynamic gas drag. We find that the dynamical implantation of CI asteroids from $\sim15$--25 au 
is very efficient with $\sim 5$\% of $\sim 100$-km planetesimals reaching stable orbits in 
the asteroid belt by the end of the protoplanetary gas disk lifetime. The efficiency is 
reduced when planetesimal ablation is accounted for. The implanted population subsequently evolved 
by collisions and was depleted by dynamical instabilities. The model can explain why CIs are 
isotopically distinct from other C-type asteroids which presumably formed at $\sim5$--10 au.   
\end{abstract}

\section{Introduction}

High-precision isotopic analyses of celestial bodies have uncovered large-scale isotopic anomalies 
that cannot be easily explained by conventional nebular or planetary processes (Dauphas \& Schauble 
2016). These anomalies affect numerous elements and demand the involvement of unusual photochemical 
or nucleosynthetic processes. Although most of these anomalies originate from nucleosynthesis, they 
cannot be attributed to a single astrophysical site.

Isotopic anomalies in meteorites, Earth, Moon, and Mars reveal patterns and clusters that have garnered 
significant interest, as they could offer insights into the processes that shaped the solar system 
in its early stages. Studies have shown that celestial bodies with isotopic anomalies can be classified 
into two main groups (Warren 2011; Budde et al. 2016; Kruijer et al. 2017; Kleine et al. 2020). 
These groups are named after the chondrites found within their respective isotopic clusters: CC 
(carbonaceous chondrite) and NC (non-carbonaceous chondrite). This categorization extends to 
differentiated meteorites and planets, including achondrites, iron meteorites, Earth, the Moon, 
and Mars.

Typically, Ivuna-type (CI) chondrites are grouped with other CC meteorites. However, iron isotopic 
anomaly measurements (expressed as $\mu ^{54}{\rm Fe}$; Schiller et al., 2020; Hopp et al. 2022) 
from CI meteorites and samples collected from the Ryugu asteroid by the Hayabusa2 mission (a Cb-type 
asteroid related to CI; Yokoyama et al. 2022) indicate that CI and Ryugu form a unique group 
(Hopp et al. 2022). This group is characterized by an Earth-like (NC-like) iron isotopic composition 
but large anomalies (CC-like) for most other elements. Recent findings on nickel isotopic anomalies 
have strengthened the case for classifying CI and Ryugu as a distinct group (Spitzer et al. 2023). 
Instead of the NC-CC dichotomy, a NC-CC-CI trichotomy is a more accurate representation (Hopp et al. 2022).

Isotopic variations documented in planetary materials most likely result from projecting of isotopic
heterogeneities that existed in the molecular cloud core parental to the solar system onto the
protoplanetary disk during collapse (Dauphas \& Schauble 2016) and/or from thermal destruction
of interstellar grains (Trinquier et al. 2009). The details on how planetary bodies acquired 
distinct isotopic compositions is very much debated (e.g., Schiller et al. 2018, Burkhardt et al. 
2019, Nanne et al. 2019, Ek et al. 2020, Lichtenberg et al. 2021, Liu et al. 2022a, Morbidelli et 
al. 2022, Izidoro et al. 2022). A comprehensive review of these contrasting perspectives is 
outside the purview of this paper. Each model offers its unique advantages and limitations, 
and no single model comprehensively explains all the observations to date. 

Burkhardt et al. (2019) and Nanne et al. (2019) argued that the NC/CC dichotomy resulted from a change 
in the isotopic composition of material accreted by the disk from the parental molecular cloud core 
(Dauphas \& Schauble 2016). This scenario was modeled quantitatively by Morbidelli et al. (2022). 
They placed the effective distance of infall, known as the centrifugal radius, at 
$r \leq 0.35$ au. This situation was shown to lead to a {\it decretion} disk with the disk material 
flowing outward for $r > 0.4$ au and expanding past the iceline at $\sim 3$--5 au in the first 0.5 Myr. 
Now, if the isotopic composition of the infalling material rapidly changed over time (in the first 
0.5 Myr), this could have plausibly produced a radial gradient because (CC-like) material accreted 
early onto the disk was transported further out than (NC-like, s-process enriched) material accreted 
later (Burkhardt et al 2019. Nanne et al. 2019). To explain the isotopic dichotomy of NC-CC {\it iron} 
meteorites, Morbidelli et al. (2022) proposed that their parent planetesimals formed (and 
differentiated) at two distinct radial locations: NC planetesimals at the silicate sublimation 
line near 1 au and CC planetesimals at the iceline near 3--5 au. 

Meteoritic evidence indicates that the NC/CC dichotomy was established early and was preserved 
until the parent-bodies of chondrites formed; several million years after solar system formation 
(Dauphas \& Chaussidon 2011 and references therein). The early formation of Jupiter (Kruijer et 
al. 2017, Nanne et al. 2019) or pressure bumps (Brasser \& Mojzsis 2020) were proposed to accomplish 
this (by halting the flow of inward-drifting CC pebbles that would otherwise pollute the inner NC 
reservoir). Liu et al. (2022a), however, pointed out a fundamental problem with these ideas: if 
Jupiter efficiently blocks drifting pebbles, the inner NC reservoir becomes quickly depleted and 
the formation of NC chondrites cannot last $\sim 1$--2 Myr. Liu et al. (2022a) proposed that the 
NC and CC reservoirs were separated by the viscous expansion of protoplanetary gas disk. They started 
with an NC-CC radial gradient by suggesting that FU Orionis--type outbursts heated up the inner disk 
and thermally destroyed (presumably) fragile CC carriers in the inner disk (Trinquier et al. 2009,
Ek et al. 2020). Then they constructed a disk with the gas inflow at $r<r^*$ and gas outflow
at $r>r^*$, where $r^* \sim 10$ au for $t < 1$ Myr, and $r^* > 10$ au for $t > 1$ Myr. They 
showed that particles starting at 1--10 au drifted into the inner disk in $\sim 1$ Myr, whereas 
those initially located beyond 20 au were delayed by the outer disk expansion, and arrived 
several Myr later. In this model, the parent bodies of ordinary and enstatite chondrites would form 
at $t \sim 1$--2 Myr and would incorporate predominantly NC material akin to achondrites, but with 
a notable contribution of CC dust. The parent bodies of carbonaceous chondrites would form at 
$t\sim 3$--4 Myr from CC dust and would be scattered into the asteroid belt by migrating Jupiter 
(Raymond \& Izidoro 2017). This model does not impose an impermeable barrier between the inner 
and outer solar system. It requires thermal modification of isotopic anomalies during the accretion 
(Onyett et al. 2023), as otherwise the isotopic compositions of some elements like Zr would not 
be reproduced (Dauphas et al. 2024).

Planetesimals forming in the vicinity of giant planets cannot survive near their original orbital 
radii. As the giant planets grow, migrate, and scatter smaller bodies, they effectively clear the 
4-35 au region of any small objects (present Jupiter and Neptune Trojans that are generally thought to be 
populations captured after the gas disk dispersal, Emery et al. 2015; but also see Pirani et al. 2019a,b). 
Nevertheless, planetesimals 
from the giant planet region can find refuge on more stable orbits within the asteroid and Kuiper 
belts (Raymond \& Izidoro 2017; Morbidelli \& Nesvorn\'y 2020). Our study investigates the 
implantation of planetesimals formed at 4-35 au into the asteroid belt, focusing on the early stages 
when the protoplanetary gas disk was still present ($\lesssim $5-10 Myr). During that time, planetesimals 
dynamically excited by the giant planets were more likely to be implanted in the asteroid belt as gas 
could circularize their highly eccentric orbits. While the implantation {\it after} gas disk dispersal is 
possible, it is inefficient (Levison et al. 2009, Vokrouhlick\'y et al. 2016), and most scattered 
planetesimals are ejected from the solar system. Outer solar system planetesimals that end up in the 
asteroid belt collide with each other and with other bodies, producing fragments (Bottke et al. 
2005, Deienno et al. 2022) believed to be the parent bodies of carbonaceous meteorites. 

Our research is closely aligned with Raymond \& Izidoro (2017) and Deienno et al. (2022) who modeled 
the early implantation of planetesimals from the giant planet region. However, we concentrate on the $\sim15$-25 au zone, as this could be the primary source reservoir for CI planetesimals (Hopp et 
al. 2022). We distribute planetesimals between 4 and 35 au, implement various growth/migration schemes 
for giant planets, consider different physical properties and life spans for the protoplanetary 
gas disk, include thermal ablation of planetesimals (Eriksson et al. 2021; Sect 3.3),
and determine the implantation efficiency of planetesimals in each case as a function 
of their starting orbital radius. The $N$-body integration method is detailed in Sect. 2, with 
results reported in Sect. 3 and discussed in Sect. 4.
    
\section{Methods}

Our simulations are set up to follow the orbital evolution of planets and planetesimals
in a protoplanetary gas disk. They are highly idealized. We adopt two different planetary 
configurations. In the first one, five planets (Jupiter, Saturn and three ice giants; one of 
which is ejected from the system after the gas disk dispersal) are placed in a compact 
resonant configuration (Case 1 from Nesvorn\'y et al. 2013). This initial configuration, with 
the 3rd ice giant between the original orbits of Saturn and Uranus, was shown to satisfy 
many solar system constraints, including the current orbits of the giant planets, capture
of Jupiter Trojans and irregular satellites, and the orbital distribution of Kuiper belt 
objects (see Nesvorn\'y 2018 for a review).
In the second one, Jupiter, Saturn, Uranus and Neptune are placed on their current 
orbits. The planets gravitationally interact with each other but are not affected by the 
gas disk (there is no realistic treatment of planet-disk interactions in our simulations). 
In some cases, we induce radial migration of planets by artificial force terms (Sect. 3.2). 

The planetesimals are given initial diameters $D=10$ or 100 km in different simulations. 
Ten thousand planetesimals are distributed between 4 and 35 au with a prescribed surface 
density $\Sigma_{\rm p} \propto 1/r$, where $r$ is the orbital radius. 
This means that we initially 
have an equal number of bodies in each semimajor axis interval. This is not intended to represent 
the real distribution of planetesimals -- which is unknown -- but gives us a good coverage of 
every possible source region. Changing the planetesimal distribution would not change the 
probability of implantation in the asteroid belt as planetesimal-planetesimal interactions 
have a negligible role compared to planet-planetesimal and gas-planetesimal interactions. 
To start with, all planetesimals have orbits with low eccentricities and low inclinations. 
The planetesimals are subject to gravitational forces of planets, do not influence each other 
or planets themselves, and feel the effects of aerodynamic gas drag.\footnote{We also include 
the effects of disk's gravitational potential on planetesimals, 
but this has not much relevance here (Vokrouhlick\'y \& Nesvorn\'y 2019).}    

The gas disk is assumed to be axisymmetric. The gas density follows a power-law distribution in 
the radial direction and has a Gaussian profile in the vertical direction:
\begin{equation}
\rho(r,z) = \rho_0 \left(r \over r_0\right)^\alpha \exp(-z^2/\zeta^2)\ ,
\label{density}
\end{equation} 
where $r$ and $z$ are the cylindrical coordinates centered at the Sun, $\rho_0$ 
is the reference gas density at $r_0=1$ au, and $\zeta$ is the scale height. 
We adopt a non-flaring disk with the fixed aspect ratio $h=\zeta/r=0.05$. The gas surface 
density is $\Sigma_{\rm g}=\sqrt{\pi} \rho_0 r_0 h r^{\alpha+1}$ in this disk model.
We set $\alpha=-3/2$ and $\Sigma_{\rm g}=2000$ g/cm$^2$ for $r=1$ au. For reference, the Minimum Mass Solar 
Nebula (MMSN; Hayashi 1981) has $\Sigma_{\rm g}\simeq1700$ g/cm$^2$ for $r=1$ au and $\alpha=-5/2$. Our 
reference disk is thus similar to MMSN but has a shallower radial profile. See Andrews (2020) for 
constraints from observations of protoplanetary disks which tend to have shallower radial profiles 
than MMSN.

Planetesimals in a gas disk experience aerodynamic drag with the drag acceleration given by
(Weidenschilling 1977)
\begin{equation}
\textbf{a}=-{3 \over 4} {C_{\rm d} \over D} {\rho(r,z) \over \rho_{\rm p}} v_{\rm r} \textbf{v}_{\rm r}\ ,
\label{drag}
\end{equation}
where $\rho(r,z)$ is the gas density from Eq. (\ref{density}), $D$ and $\rho_{\rm p}$ are the 
diameter and bulk density of planetesimals, $\textbf{v}_{\rm r}$ is the relative velocity of planetesimals 
with respect to gas (the orbital motion of gas is slightly sub-Keplerian due to radial pressure
support; Adachi 1976), and $C_{\rm d}$ is the aerodynamic drag coefficient.
In the highly supersonic regime, which is relevant when planetesimals are scattered to high-eccentricity orbits 
by planets, $C_{\rm d}$ approaches 2. We implement the full dependence of $C_{\rm d}$ on the Mach and 
Reynolds numbers from Brasser et al. (2007).  The planetesimals are given $\rho_{\rm p}=1$ g/cm$^3$ 
(a fiducial value for outer solar system bodies).

The integrations were performed with the \textit{Swift} code (Levison \& Duncan 1994), an efficient 
$N$-body code based on the Wisdom-Holman map (Wisdom \& Holman 1991), which is capable of simulating
close encounters of planetesimals to planets. Gas drag was implemented in the kick part of the 
integrator that deals with massless bodies (planetesimals); drag is not applied to planets. We 
follow the orbits of planets and planetesimals for the assumed duration of the gas disk and check 
on how planetesimals become re-distributed by the combined effect of planet scattering and gas drag. 
The gas density is either assumed to be unchanging with time or we decrease it following an exponential 
decay with one e-fold $\tau_{\rm g}$. The evolution of planetesimals after the gas disk dispersal is 
ignored in the present simulations, but we discuss it in Sect. 4. 

\section{Results}

\subsection{Base model}

In our base model, Jupiter and Saturn start in the 3:2 resonance with Jupiter at 5.6 au and Saturn 
at 7.5 au (e.g., Morbidelli et al. 2007).  Raymond \& Izidoro (2017) modeled the runaway gas accretion
of Jupiter and Saturn and showed that different accretion timescales ($\sim 10^4$--$10^6$ yr) yield 
similar results. We therefore ignore this issue and introduce Jupiter and Saturn immediately at 
the beginning of our simulations ($t=0$). In reality, the gas giants accreted their envelopes
with some delay, $\Delta t$, after the condensation of the first solar system solids 4.568 Gyr 
ago (see Kleine et al. 2009 for a review). Our simulations thus effectively start $\Delta t$
into the protosolar nebula evolution. 

All planetesimals are given initial diameter $D=100$ km, because 
this is the characteristic size of planetesimals in the Solar System (e.g., Morbidelli et al. 
2009), and are introduced at $t=0$. Planetesimal ablation is ignored (see Sect. 3.3).
We start with the protoplanetary gas disk described in Sect. 2 
($\alpha=-3/2$ and $\Sigma_{\rm g}=2000$ g/cm$^2$ for $r=1$ au) and exponentially decrease the gas 
density with $\tau_{\rm g}=3$ Myr. The simulation is run over 10 Myr. This is not meant to represent
the real duration of the protosolar gas disk. Observations suggest that protoplanetary gas disks
last $\sim 2$--10 Myr (Haisch et al. 2001, Williams \& Cieza 2011). In the solar system, we have no 
evidence from meteorites for primary asteroid accretion after $\simeq 4$--5 Myr and this is often 
interpreted as marking the end of the protosolar gas disk (Dauphas \& Chaussidon 2011). The results 
reported below for $t=4$ Myr should thus be the most relevant (assuming that $\Delta t \lesssim 
1$ Myr). 

The slow gas decay with $\tau_{\rm g}=3$ Myr and abrupt termination of the disk at $t=4$ Myr would 
schematically correspond to slow viscous evolution of the disk terminated by its rapid photoevaporation. 
Faster initial gas disk removal by MHD winds is also tested (e.g., Komaki et al. 2023; see below). 
Three ice giants are introduced in the simulation at $t=3$ Myr. The relatively late formation of 
ice giants could explain why they did not accrete more massive gas envelopes (Pollack et al. 1996).
They are placed in a resonant chain with the inner ice giant in the 3:2 resonance with Saturn, Uranus in the 2:1 
resonance with the inner gas giant, and Neptune in the 3:2 resonance with Uranus (Sect. 2; Nesvorn\'y 
et al. 2013, Deienno et al. 2017). 

\begin{figure}
\epsscale{0.6} 
\plotone{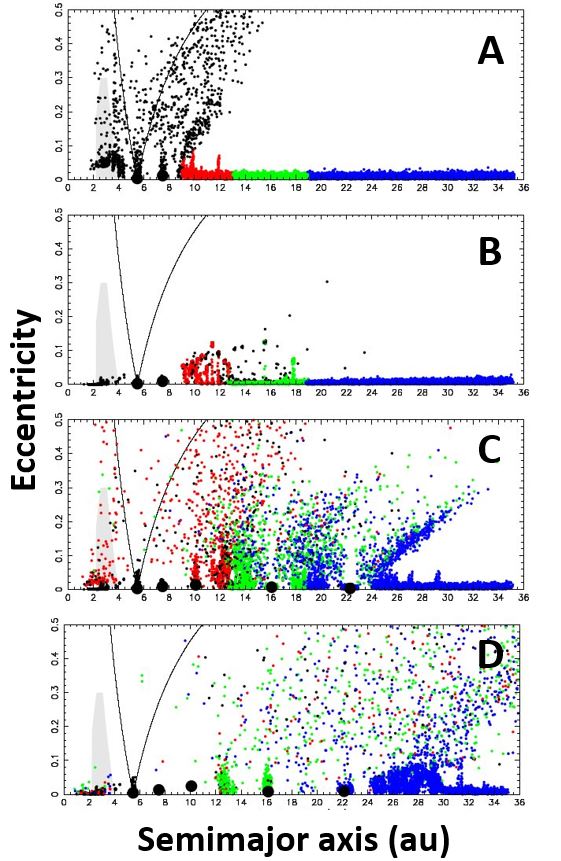}
\caption{Dynamical evolution of outer solar system planetesimals. All planetesimals were given
diameter $D=100$ km and they were assumed to be there at the beginning of the simulation. The starting 
radial distance of planetesimals is color coded (black for 4--9 au, red for 9--13 au, green for 13-19 au, 
and blue for 19--35 au). Initially, there is an equal number of planetesimals at each orbital radius. 
The panels show different stages from our base model simulation: (a) shortly after Jupiter and Saturn
were placed in the disk (time $t=10^4$ yr), (b) just before the ice giants were introduced ($t=3$ Myr),
(c) shortly after the ice giants were introduced ($t=3.1$ Myr), and (d) at the end of the 
simulation ($t=10$ Myr). The five planets remain in a compact resonant configuration (Nesvorn\'y et 
al. 2013; planet migration is ignored). The solid black lines are Jupiter-crossing orbits 
with $q=a(1-e)=5.2$ au and $Q=a(1+e)=5.2$ au. The shaded area at 2.2--3.5 au approximately outlines 
the orbital region of the current asteroid belt.}
\label{expon}
\end{figure}

During the initial stage, Jupiter and Saturn scatter planetesimals from 4--9 au (Fig. \ref{expon}). 
Many planetesimals are scattered inward, where their orbits can be circularized by gas drag to end 
up in the asteroid belt, while others are scattered outward and often end up being ejected from the 
solar system. We define the implantation probability from the initial orbital distance $r$ as
\begin{equation}
P_{\rm implant}(r)= { N_{\rm implant}(r) \over N_0(r) }\ ,  
\label{implant2}
\end{equation}
where $N_0(r)$ is the initial number of planetesimals in a small interval $\Delta r$ around 
$r$ and $N_{\rm implant}(r)$ is the number of planetesimals starting in $\Delta r$ around $r$ 
{\it and} ending in the asteroid belt. By monitoring $N_{\rm implant}(r)$ as a function 
of time in our simulations and reporting $P_{\rm implant}(r,t)$ at time $t$, we can test different 
scenarios for the gas disk dispersal (e.g., $P_{\rm implant}(r,t=4\, {\rm Myr})$ corresponds
to gas disk dispersal at $\Delta t + 4$ Myr). 

Figure \ref{implant} shows that the implantation
probability of planetesimals from 4--9 au can be as high as 50\%, at least for the base simulation 
setup adopted here. By $t=3$ Myr, the 4--9 au region is practically empty (except for a small 
population Jupiter co-orbitals). The next stage starts when the ice giants are introduced in 
the disk and scatter planetesimals from 9--24 au. The final implantation probabilities from 9--24 au 
are generally lower than those from the Jupiter/Saturn zone but still relatively high (1--20\%; Fig. 
\ref{implant}). There is a narrow region at 12--14 au where some planetesimals survive on low-eccentricity 
orbits between the inner ice giant and Uranus (bottom panels in Fig. \ref{expon}). The outer disk at 
$r>24$ au, which is thought to be the main source of Kuiper belt objects and comets, survives undisturbed 
(additional bodies are injected to it from $r<24$ au). The implantation probability from $r>26$ au to 
the asteroid belt is practically zero. The outer disk planetesimals can be implanted into the asteroid 
belt after the gas disk dispersal ($\sim 10^{-6}$ capture probability; Levison et al. 2009, 
Vokrouhlick\'y et al. 2016).
  
\begin{figure}
\epsscale{0.6}
\plotone{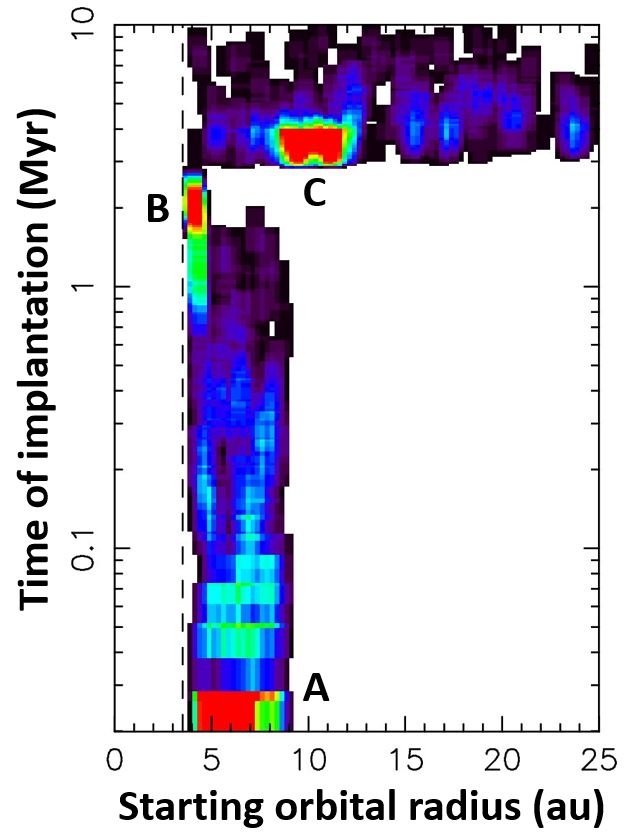}
\caption{Time of implantation of outer solar system planetesimals into the asteroid belt. Warmer colors 
indicate higher implantation probabilities. The probabilities shown stand for $P_{\rm implant}(r)$ 
calculated from Eq. (\ref{implant2}) but we also factor in the time when bodies are implanted in the 
belt $t_{\rm implant}$ and show $P_{\rm implant}(r,t=10\,{\rm Myr})$ as a function of $t_{\rm implant}$. 
The binning is done in $\log(t)$ thus highlighting the implantation sources at late times. The three red 
regions are: (\textbf{A}) early implantation from the Jupiter/Saturn reservoir, (\textbf{B}) planetesimals 
implanted by gas-drag migration, and (\textbf{C}) late implantation from the ice giant reservoir. We 
show the results for the whole simulation time span (10 Myr).}
\label{color}
\end{figure}

Overall, we identify three implantation stages. During the early stage, Jupiter and 
Saturn form and scatter planetesimals from 4--9 au (implantation stage A in Fig. \ref{color}).
This stage is relatively brief ($<10^5$ years after Jupiter and Saturn reach their final masses). 
Interestingly, some planetesimals starting at $r \simeq 4$ au migrate on low-eccentricity orbits 
into the asteroid belt region. This is helped by Jupiter that excites the orbital eccentricity 
of planetesimals, which then feel a stronger headwind and experience faster orbital decay.
They reach $r<3.5$ au in 1.5--2.5 Myr (stage B in Fig. \ref{color}). 
The third stage of implantation starts when the ice giants are introduced in the simulation and 
lasts for $\simeq 1$ Myr. During this stage (stage C in Fig. \ref{color}), many planetesimals from 
9--24 au are scattered inward by the ice giants and end up in the asteroid belt. The planetesimals 
starting at 9--12 au have 10--20\% implantation probabilities, and those starting at 15--24 au 
have 1--3\% implantation probabilities (both reported here for $t=4$ Myr). The implantation probability 
drops well below 1\% near 13--14 au, where many bodies manage to survive on low-eccentricity orbits
(Fig. \ref{expon}; these bodies would be removed -- practically all being ejected from the 
solar system -- during planet migration/instability after the gas disk dispersal). 

The great majority of planetesimals implanted in the asteroid belt end up on low-eccentricity and 
low-inclinations orbits. There is a correlation between the time of implantation and the orbital 
radius of implanted planetesimals. The planetesimals scattered by the giant planets early tend to 
end up in the outer asteroid belt with $a \simeq 3$ au. This happens because the gas density is 
still relatively high during the early stages and orbits circularize by gas drag before they can 
reach smaller orbital radii. The planetesimals that are scattered late can end up in the inner part
of the asteroid belt (2--2.5 au) or even in the terrestrial planet region near 1 au.
This happens because the gas density of an aging disk is relatively low and Jupiter is capable of
scattering planetesimals to very small orbital radii before gas drag becomes important. For the 
model setup that we explore here, with Jupiter and Saturn forming early and the ice giants forming 
late, the correlation discussed above leads to a trend where the planetesimals from the Jupiter/Saturn 
zone preferentially populate the outer part of the asteroid belt and the planetesimals from the ice 
giant zone become more widely distributed across the whole asteroid belt.  

\subsection{Auxiliary simulations} 
    
We performed many additional simulations. The simulations with four giant planets on their current
orbits produce results similar to the base model. The main difference is that the final implantation 
probability from the Jupiter/Saturn region is significantly lower ($\sim 1$\% compared to $>10$\% 
for the compact five planet system). This happens because the current orbit of Jupiter is somewhat 
eccentric (mean eccentricity 0.044), this forces orbital oscillations of orbits and enhances 
the gas drag on bodies implanted in the asteroid belt (the $v_{\rm r}$ term in Eq. (\ref{drag}) is 
larger). Consequently, the planetesimals that are implanted in the asteroid belt early in the 
simulation manage to drift below 2 au before the gas disk dispersal. We do not consider this case
to be particularly realistic because, in reality, Jupiter's eccentricity would have been damped 
by disk torques. The implantation probability from the 15--25 au region is lower in the four 
planet case (0.2--1\% for $t=4$ Myr) than in the five planet case (1-3\% for $t=4$ Myr) because, 
in addition to the effect described above, many planetesimals survive in the gaps between planets 
(there are large radial gaps between planets in our four planet case with planets near their 
current locations).        

The simulations with migrating planets show additional dynamical effects, mainly related to
capture of planetesimals in orbital resonances. The relevance of these effects for the implantation 
of planetesimals in the asteroid belt, however, seems to be modest. The only migration setup that 
yielded results substantially different relative to that with the static planets is the case where 
Jupiter was assumed to form at large orbital radius ($r>10$ au) and migrate inward (Deienno et 
al. 2022). There are two competing effects in these simulations. On one hand, migrating Jupiter 
has an effectively larger feeding zone and can scatter a larger population of outer solar 
system planetesimals. This would increase the population of implanted planetesimals (Deienno et 
al. 2022). On the other hand, Jupiter's migration would delay the whole process: with Jupiter
at $>10$ au, the early-scattered planetesimals do not reach the asteroid belt and end up at $>5$ au. 
This leads to additional considerations. For example, if the gas density at $<5$ au decreased 
over time and was already orders of magnitude below MMSN when Jupiter reached its current 
orbital radius, the overall implantation efficiency into the asteroid belt could become less 
efficient. 

\begin{figure}
\epsscale{0.9}
\plotone{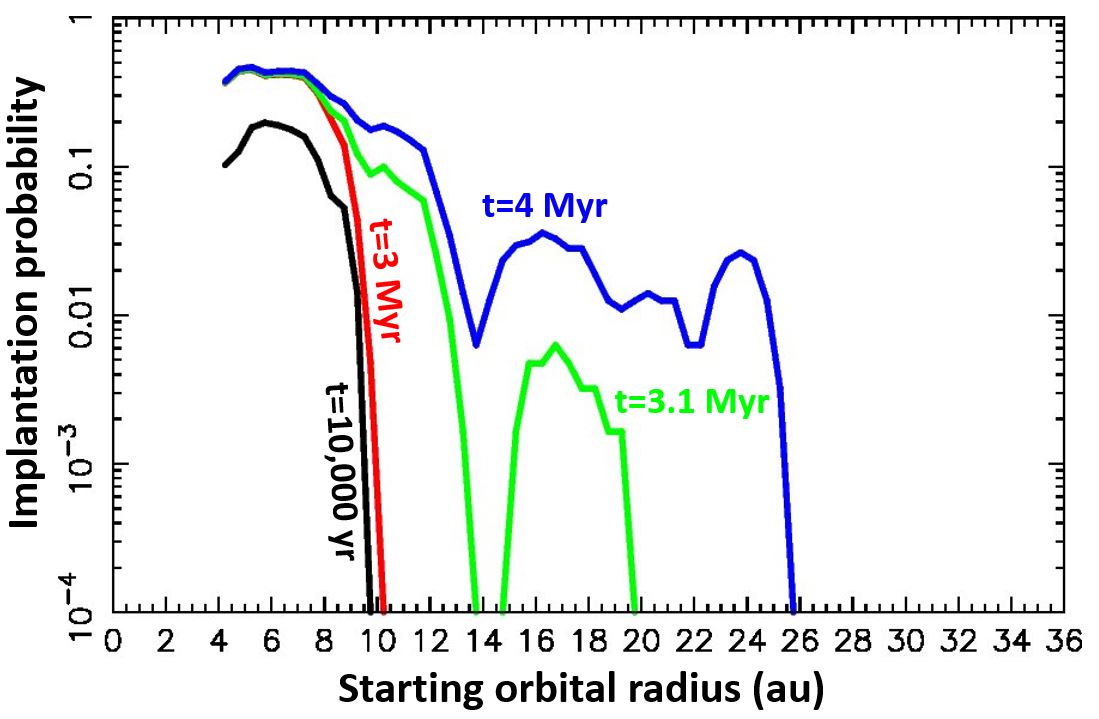}
\caption{The implantation probability of outer solar system planetesimals into the asteroid belt.
The lines show the implantation probability computed at different times (black for $t=10^4$ yr, red 
for $t=3$ Myr, green for $t=3.1$ Myr, blue for $t=4$ Myr). To obtain the number of planetesimals 
implanted into the asteroid belt from each orbital radius, the final profile shown here (blue line)
would have to be multiplied by the number of planetesimals originally available at each orbital 
radius. The final implantation probability is roughly 10--30 times lower for $r=15$--25 au than for
$r=5$--10 au. For comparison, Cb-type asteroids like Ryugu represent $\sim 10$\% of all C-type 
asteroids (Rivkin 2012, DeMeo et al. 2009).}
\label{implant}
\end{figure}

\begin{figure}
\epsscale{0.9} 
\plotone{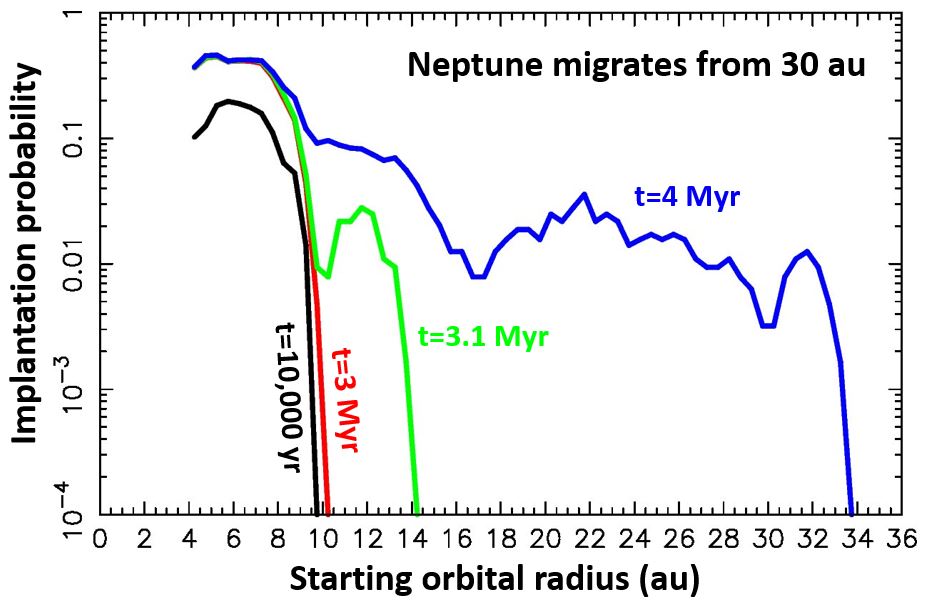}
\caption{The implantation probability of outer solar system planetesimals into the asteroid belt.
Unlike in Fig. \ref{implant}, where the outer planets were held at constant orbital radii, here
the ice giants were assumed to radially migrate. This extends the radial range from which planetesimals 
can be implanted into the asteroid belt ($4<r<35$ au here vs. $4<r<26$ au in Fig. \ref{implant}).  
The lines show the implantation probability computed at different times (black for $t=10^4$ yr, red 
for $t=3$ Myr, green for $t=3.1$ Myr, blue for $t=4$ Myr). To obtain the number of planetesimals 
implanted into the asteroid belt from each orbital radius, the final profile shown here (blue line)
would have to be multiplied by the number of planetesimals originally available at each orbital 
radius.}
\label{migra}
\end{figure}

If, as we argue here, the parent bodies of CI meteorites formed at $r \gtrsim 15$ au, another 
process that may be relevant here is the formation and orbital migration of ice giants. Above 
we assumed that the ice giants formed at $t=3$ Myr, and placed them directly into a resonant 
chain expected from the convergent gas-driven migration. Here we account for their orbital 
migration. Uranus and Neptune were placed at 20 and 30 au, respectively, and migrated inward on a 
timescale consistent with the Type-I torques (Paardekooper et al. 2011). After several attempts, the 
eccentricity damping timescale was found such that the ice giants ended up in practically the 
same orbital configuration as the one used for the static simulations (Uranus and Neptune 
at 16 and 22 au). Figure \ref{migra} shows the implantation probability in this case. Compared       
to Figure \ref{implant}, the main difference is that planetesimals can be implanted into the
asteroid belt from a wider range of initial orbital radii. The implantation probability from 
$r=25$--32 au is $\sim 1$--2\%. As a side effect of Neptune's migration from 30 to 22 au, the 
outer planetesimal belt is destroyed. It may therefore be more difficult to understand the 
Kuiper belt formation in this case (Nesvorn\'y 2018; unless the outer disk planetesimals 
accreted after Neptune reached $r<25$ au).  

The simulations with 0.1, 0.01 and 0.001 MMSN, where the gas density was decreased by 10, 100 and 
1000 times relative to MMSN, show lower implantation probabilities. For example, for 0.01 MMSN and 
the compact five planet system, the implantation probabilities from 10-25 au to the asteroid belt 
are only 0.1-1\%. The lower gas density in these cases reduces the effects of gas drag and 
planetesimals are often scattered past the asteroid belt into the terrestrial planet 
region.\footnote{The implantation process is a competition between planet scattering, which leads 
to orbits with lower semimajor axes and higher eccentricities, and gas drag, which acts to 
circularize orbits. If the gas drag is reduced, planet scattering wins and the orbits end up 
being scattered closer to the Sun.} More realistic gas disk profiles obtained from hydrodynamical 
simulations of viscous disks also show reduced gas densities (e.g., Morbidelli \& Crida 2007). 
Inviscous disks with disk winds indicate lower gas densities for $r<5$ au as well (e.g., Suzuki 
et al. 2010).    

Additional simulations were performed for different planetesimal sizes and disk life times.
The results are generally consistent with those obtained in Raymond \& Izidoro (2017) and we 
do not discuss them in detail here. As the gas drag scales with $\rho/D$ (Eq. \ref{drag}), 
where $\rho$ is the gas density and $D$ is the planetesimal diameter, the results for 
larger (smaller) planetesimals are equivalent to those obtained for lower (higher) gas 
densities. As for the disk life time, some insights can be gleaned from Figs. \ref{expon}--\ref{color}.
If, for example, the gas disk would be removed shortly after the formation of ice giants 
($t=3$ Myr in our base simulation), the implantation probability from 10-25 au would be strongly 
reduced. The bulk of implanted CI planetesimals would be sourced from 15-20 au in this case 
(0.2-0.7\% probability; the green line in Fig. \ref{implant}). In addition, as the implantation 
probability drops to $<10^{-4}$ near 14 au, CCs and CIs would be sampling radially separated 
reservoirs ($r<12$ au for CCs and $15<r<20$ au for CIs), and this could help to explain their 
distinct isotopic properties (see below for a discussion of this issue). 

\subsection{Planetesimal ablation}

We follow the model of Eriksson et al. (2021) to account for thermal ablation of planetesimals. 
The outer solar system bodies are assumed to have formed with a substantial fraction of H$_2$O
ice that is well mixed with the silicate and carbon grains. Thus, if H$_2$O sublimates on body's
surface, the silicate and carbon grains would be carried away. This ignores the possibility that 
the H$_2$O ice sublimation could leave behind a refractory crust.\footnote{Eriksson et al. (2021) 
argued that any refractory crust would be blown away by the vapor pressure buildup underneath 
the surface. It is not clear, however, whether this process would remove the whole crust as 
vapor could diffuse outward and escape (e.g., Sch\"orghofer \& Hsieh 2018), leaving the refractory 
crust in place.} Other volatile species such as CO$_2$ or CO are not considered here.  
  
The mass loss of a planetesimal due to ablation is given by
\begin{equation} 
\dot{m} = -4 \pi R^2_{\rm pl} P_{\rm sat}(T_{\rm pl}) \sqrt{\mu \over 2 \pi R_{\rm g} T_{\rm pl} }\ ,
\end{equation}
where $R_{\rm pl}=D/2$ is the planetesimal radius, $P_{\rm sat}$ is the saturated vapor pressure of H$_2$O,
$T_{\rm pl}$ is the surface temperature of the planetesimal, $\mu=0.0180153$ kg mol$^{-1}$ is the 
molecular weight of H$_2$O, $R_{\rm g}=8.3145$ J mol$^{-1}$ K$^{-1}$ is the ideal gas constant
(D'Angelo \& Podolak 2015). The saturated vapor pressure of H$_2$O is calculated from the 
polynomial expressions given in Fray \& Schmitt (2009). The equilibrium surface temperature
of planetesimal surface, $T_{\rm pl}$, is computed from (Ronnet \& Johansen 2020)
\begin{equation}
\sigma_{\rm SB} (T^4_{\rm pl} - T^4)  = {C_{\rm d} \rho v_{\rm rel}^3 \over 32} - {\dot{m} L \over 4 
\pi R^2_{\rm pl}}\ ,
\label{equilib}
\end{equation} 
where the gas density $\rho$ is given in Eq. (\ref{density}), $v_{\rm rel}$ is the relative velocity
of planetesimals with respect to gas, $\sigma_{\rm SB}=5.670374419 \times 10^{-8}$ W m$^{-2}$ K$^{-4}$
is the Stefan-Boltzmann constant, and $L=2.8 \times$ J kg$^{-1}$ is the latent heat of H$_2$0 
vaporization given here for low temperatures and pressures (Eriksson et al. 2021). The midplane
temperature of the disk was approximated from $T=T_{\rm 1au}(r/{\rm au})^{-3/7}$, where 
$T_{\rm 1au}=150$ K is the temperature at 1 au (Chiang \& Goldreich 1997). The first term
on the right-hand side of Eq. (\ref{equilib}) stands for heating due to gas friction, the 
second term is cooling due to the ice sublimation. 

\begin{figure}
\epsscale{0.7} 
\plotone{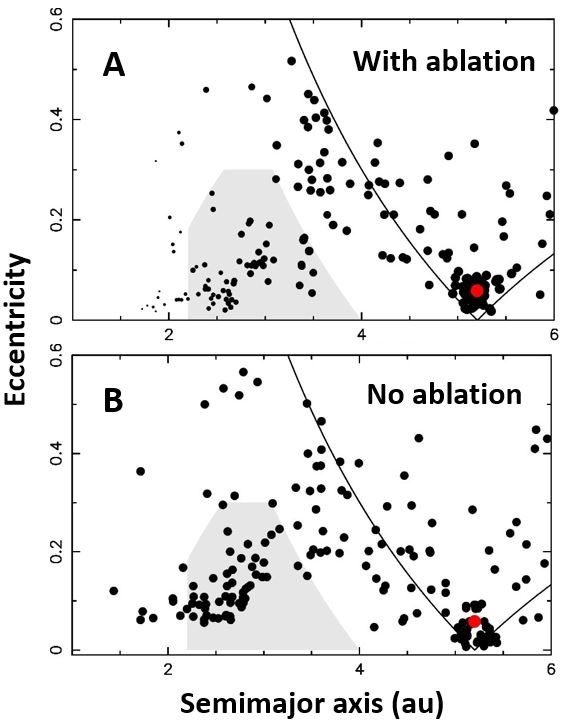}
\caption{The implantation of outer solar system planetesimals in the asteroid belt for the case with 
(panel A) and without (panel B) planetesimal ablation. The symbol size is proportional to 
the planetesimal diameter. All planetesimals start with $D=100$ km. The smallest planetesimals
shown in panel A have $D \simeq 10$ km. In both cases, we show the situation shortly after the start 
of simulations ($t=10^5$ yr).}
\label{ablate}
\end{figure}

We implemented planetesimal ablation in \textit{Swift} and performed several simulations to test 
the effects of ablation in different cases. Figure \ref{ablate} compares the cases with and 
without planetesimal ablation in our base model (initial $D=100$ km, $\alpha=-3/2$, 
$\Sigma_{\rm g}=2000$ g/cm$^2$ for $r=1$ au, $\tau_{\rm g}=3$ Myr). With planetesimal ablation,
the bodies scattered on orbits with larger eccentricities are more heated and lose more mass. 
This produces a clear trend where planetesimals implanted in the inner main belt ($2<a<2.5$ au)
are smaller than planetesimals implanted in the outer main belt ($3<a<3.5$ au). The 
characteristic size of planetesimals implanted at different radii is: $D \simeq 30$ km for $a 
\simeq 2$ au, $D \simeq 50$ km for $a \simeq 2.5$ au, and $D \simeq 80$ km for $a \simeq 3$ au. 

Taken at face value, this would imply that a similar trend should exist among main-belt 
asteroids. We checked on that by plotting the size distribution of dark main-belt asteroids 
(albedo $p_{\rm V}<0.1$) from the WISE data (Mainzer et al. 2011) but could not find any clear 
trend. Things can become more complicated when one considers the initial size distribution of 
planetesimals. Ribeiro de Sousa et al. (2023) studied the effects of ablation on the size 
distribution of outer solar system planetesimals implanted in the asteroid belt and found that the size 
distribution generally evolved to become steeper. This most likely happens because large planetesimals 
can be scattered to high orbital eccentricities and ablate faster that small ones. A detail 
comparison of the ablation model with the size distribution of asteroids is left for future work.

\begin{figure}
\epsscale{0.9} 
\plotone{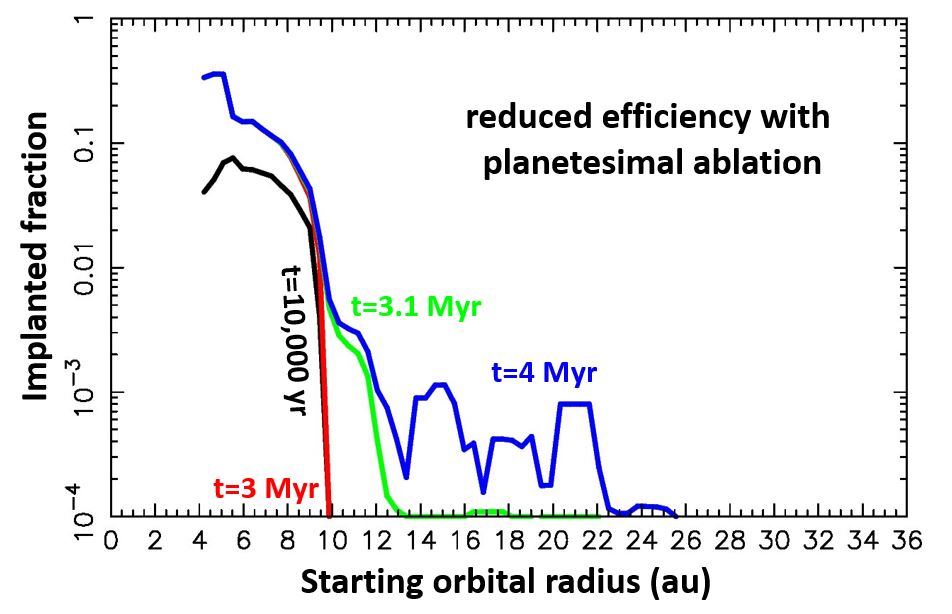}
\caption{The implanted fraction of outer solar system planetesimals into the asteroid belt.
Unlike in Fig. \ref{implant}, here we account for planetesimal ablation (Eriksson et al. 2021).
The implanted fraction is the volume fraction of planetesimals starting at a particular orbital 
radius that ends up in the asteroid belt. This accounts for both the implantation probability
of individual objects and their mass loss due to ablation. The lines show the implanted fraction
at different times (black for $t=10^4$ yr, red for $t=3$ Myr, green for $t=3.1$ Myr, blue for 
$t=4$ Myr).}
\label{fract}
\end{figure}

Planetesimal ablation acts to reduce the implantation efficiency. We quantify that by computing 
the volumetric fraction of planetesimals that end up in the asteroid belt (Fig. \ref{fract}). 
The implanted fraction combines the implantation probability with the mass loss due to ablation.
We find that the implanted fraction from different initial orbital radii is characteristically
a factor of $\gtrsim 10$ lower when planetesimal ablation is accounted for. We moreover find, with
ablation, that the implanted fraction decreases during the late stages of evolution when the gas 
disk density decreases. This happens as lower gas densities imply lower aerodynamic drag on 
planetesimals; bodies can reach very high-eccentricity orbits, heat up, and rapidly ablate. 
Planetesimal ablation could thus help to reconcile the relatively small mass of the asteroid belt 
with our expectations from the implantation modeling (this problem is discussed in Sect. 4; 
also see Deienno et al. 2022).

\section{Discussion}

Our results show that it is possible to implant planetesimals into the asteroid belt even 
from very distant orbits (also see Raymond \& Izidoro 2017). If anything, this process is too 
efficient, which makes it highly likely that it played a significant role in populating the 
main asteroid belt with volatile and organic-rich planetesimals. As discussed above and 
illustrated in Fig. \ref{color}, we identify three episodes of planetesimal implantation: 
{\bf A.} Planetesimals originally at 4 to 9 au are very rapidly excited and implanted by 
the growth of Jupiter and Saturn. {\bf B.} Approximately 2 Myr later, the outer solar system 
planetesimals from 4--5 au make their way into the asteroid belt. These planetesimals have 
slowly drifted from their original orbits by gas drag (this is helped by small orbital 
eccentricities of planetesimals induced by Jupiter; the drift would be reduced for lower 
forced eccentricities or lower gas densities). These planetesimals come from the same source 
region as the first wave and are expected to have the same chemical and isotopic composition. 
{\bf C.} A second wave of planetesimals are implanted in the main belt by the growth of the 
ice giant planets ($t \simeq 3$--4 Myr in our base simulation). They primarily come from 
9--14 au and include planetesimals coming from as far as $\sim 25$ au. 

It is conceivable that temporal changes in the composition of infalling material and/or thermal
processing created an isotopic gradient in the outer solar system.\footnote{The isotopic divide 
between CC- and CI materials, so far, was only identified for Fe and Ni isotopic anomalies 
(Hopp et al. 2022, Spitzer et al. 2023). A possible explanation is that the isotopic heterogeneity of Fe and Ni in the outer 
solar system was dominated by a different carrier phase (e.g., non-refractory Fe,Ni metal) 
compared to other elements that display isotopic anomalies, e.g. Ti and Cr.}
As discussed by Hopp et al. (2022), the discrete excitation of planetesimals by the giant planets 
could explain why two distinct populations of carbonaceous asteroids seem to exist in the 
asteroid belt, namely CC and CI. The first planetesimals implanted in the main belt would come 
from 4--9 au and we identify those with CC. The planetesimals implanted by the ice giants come 
primarily from beyond 9 au, and this would include CIs.

The dichotomy between CCs and CIs could reflect isotopically distinct source ring reservoirs 
(e.g., Izidoro et al. 2022) or the nature of the implantation process. For example,  
asteroids with intermediate compositions between CI and CC could exist, originating from 
12--16 au, but they would be rare as they would have been less affected by the giant planets 
while nebular gas was still around (Fig. \ref{implant}), and would have been cleared from the disk by 
planetesimal-driven planet migration after gas dissipation. The likelihood of these planetesimals 
to end in the asteroid belt is very low (Vokrouhlick\'y et al. 2016); they would not be represented 
in the meteorite collection. The CC-CI divide could thus be due to discrete excitation by the 
giant planets in relationship to gas dissipation. 

Excitation of planetesimals by the ice giants would have launched a fraction of them inwards 
to be implanted in the main belt, but the majority would have been launched outwards 
and some would end up in the Oort cloud (Dones et al. 2015). CI and Ryugu could share the same 
heritage as Oort cloud comet. The physical properties of CI and Ryugu are clearly distinct 
from Oort cloud comets, starting with the fraction of ice that they contain. Asteroids larger 
than $\sim 10$ km in size should have retained their ice for the lifetime of the solar system 
(Sch\"orghofer \& Hsieh 2018), so the difference in the ice-to-rock ratio between CI/Ryugu and 
comets must have been established early. A plausible explanation is that it could be due to 
ice sublimation in planetesimals before their orbits were stabilized in the asteroid belt 
(Sect 3.3). Impact energy released during collision in the main belt could have also played a 
role, as observation of Ryugu and Bennu show that they have a rubble-pile structure likely 
produced by impacts. 


The excitation-implantation process highlighted above could by itself explain the isotopic 
divide documented between CI and CC. This selective excitation mechanism could have acted on 
more clustered populations of carbonaceous planetesimals than the smooth gradient envisioned 
above, which would help create distinct isotopic populations. Izidoro et al. (2022) suggested 
that outer solar system planetesimals formed in distinct rings at the orbital radii corresponding 
to the sublimation lines of different volatile species (H$_2$O, CO, etc.). Thus, even if 
the implantation works equally well for a continuous range of orbital radii, the implanted 
planetesimals would fall into distinct isotopic groups; they would reflect the isotopic 
composition of the ring reservoirs.  

While implantation of outer solar system planetesimals by the giant planets can potentially 
explain the presence of multiple groups of carbonaceous asteroids in the main belt, notably 
CC and CI isotopic families, there are two major issues that need to be considered with 
this scenario: ({\it i}) the implantation efficiencies are very high, and ({\it ii}) the 
implanted planetesimals end up on near circular and near co-planar orbits in the asteroid 
belt.  We discuss these issues below.

Assuming that the mass of planetesimals that remained in the giant planet region 
after the giant planets formed was comparable to that incorporated in planet's cores, say $\sim 10$ 
$M_{\rm Earth}$, several Earth masses in planetesimals would be implanted into 
the asteroid belt. For comparison, the total mass of asteroids is only $\sim 5 \times 10^{-4}$ 
$M_{\rm Earth}$ (DeMeo \& Carry 2013). Dynamical depletion of the asteroid belt after the gas disk
dispersal would help to reduce the mass but clearly not enough to wipe out four orders of magnitude 
(the main belt is estimated to loose $\sim 80$-95\% of its mass over 4.5 Gyr; see Raymond \& 
Nesvorn\'y 2022 for a review). Collision grinding could also help to reduce the implanted mass 
(Bottke et al. 2005, Deienno et al. 2022).
 
The simplest solution to this problem would be if very few planetesimals remained in the giant 
planet region when the accretion of giant planets was reaching completion. Levison et al. (2015) 
studied growth of the giant planets from accreting pebbles and planetesimals, and found that 
$\gtrsim 1$ $M_{\rm Earth}$ in planetesimals remained in the giant planet region when the planets 
formed. When combined with the high implantation efficiency, such mass would be excessive. Perhaps 
giant planet formation is not well understood. It is also not clear how the low mass in 
planetesimals in the giant planet region could be reconciled with studies of the Kuiper belt 
that require 15--20 $M_{\rm Earth}$ in planetesimals at $\sim 20$--30 au (see Nesvorn\'y 2018 for
a review). 


The implantation efficiency of outer solar system planetesimals in the asteroid belt drops when 
we adopt low gas densities (Sect. 3). When the gas density is low, planetesimals are preferentially 
scattered to $r<2$ au (without ablation; see below), where they would contribute to the formation of 
the terrestrial planets. To test this, we modified the nominal setup of simulations described in the 
Sect. 3: the initial gas density was decreased to 1/10 of MMSN and $\tau_{\rm g}=1$ Myr. With this 
setup, the implantation probabilities from 4--9 au and 15--20 au are $\simeq 2$--10\% and 0.1--0.5\%,
respectively, a factor of $\sim 10$ below the nominal probabilities obtained in Sect. 3. 
The low gas densities at $<5$ au appear most plausible in the disk wind models (e.g., Suzuki et al. 
2010, Komaki et al. 2023) or if the planetesimal scattering started late (Raymond \& Izidoro 2017).
D'Angelo et al. (2021) presented growth models of Jupiter where it takes over 3 Myr for Jupiter
to start the runaway gas accretion.
 
  
Cedenblad et al. (2021) studied erosion of planetesimals by gas in a protoplanetary disk and found that 
planetesimals on eccentric orbits rapidly erode. The erosion timescale can be extremely short 
($\sim 100$ yr for eccentric orbits in the inner disk). It is therefore also possible that planetesimals 
scattered by the giant planets eroded before being able to reach the low-eccentricity orbits 
in the asteroid belt. In addition, the effects of erosion depend on planetesimal composition and
this would presumably produce a trend with planetesimal's initial orbital distance. In Sect. 3.3,
we studied the effects of planetesimal ablation (Eriksson et al. 2021). We found that the volumetric 
fraction of implanted planetesimals is strongly reduced when the effects of ablation are accounted for. 
Taken together, it thus seems possible that lower gas densities, planetesimal erosion and ablation, 
and/or reduced initial mass in planetesimals in the giant planet region could resolve problem ({\it i}). 

As for problem ({\it ii}), bodies implanted in the asteroid belt end up on orbits with low eccentricities and 
low inclinations. The orbits of current asteroids, however, are dynamically excited with eccentricities 
0--0.3 and inclinations 0--20$^\circ$. The effects of the giant planet migration/instability 
after the gas disk dispersal may not, in general, be strong enough to sufficiently excite orbits 
(Roig \& Nesvorn\'y 2015, Brasil et al. 2016) but Deienno et al. (2018) identified at least one 
instability case with strong excitation. It is also possible that the asteroid belt was excited (and 
depleted) during the dispersal of the protoplanetary disk when the secular resonances with planets 
were on the move (Ward 1981, Liu et al. 2022b). The timing inferred from the cooling rates of iron 
meteorite parent bodies indicates that disruptive collisions in the asteroid belt persisted at 
$\sim 10$ Myr after CAIs (Matthes et al. 2018, Hunt et al. 2022). 

\section{Conclusions}

Our simulations show that planetesimals from a large range of heliocentric distances could have been 
implanted into the asteroid belt during the growth and migration of giant planets in a protoplanetary gas 
disk. Most outer solar system planetesimals implanted in the asteroid belt originate in the formation 
region of Jupiter/Saturn (5--10 au), but some would come from further away (10-25 au). In this context, 
the peculiar Fe isotopic heritage and primitive chemical characteristics of CI chondrites and Ryugu 
(Yokoyama et al. 2022, Hopp et al. 2022) could be explained if they formed in the vicinity of the 
birthplaces of Uranus and Neptune. 

\acknowledgements

This work benefited from discussions with Tetsuya Yokoyama and and Hayabusa2 team members. 
The work of D.N. was supported by the NASA Emerging Worlds program. The work of D.V. was supported 
by the Czech Science Foundation (grant number 21--11058S). R.D. acknowledges support from the NASA 
Emerging Worlds program, grant 80NSSC21K0387. The work of ND was supported by NASA grants 
80NSSC20K1409 (Habitable Worlds), 359NNX17AE86G (LARS), NNX17AE87G and 80NSSC20K0821
(Emerging Worlds), EAR-2001098 (CSEDI), and a DOE grant. We thank the anonymous reviewer for 
helpful suggestions.

\clearpage

\clearpage

\clearpage

\clearpage

\clearpage

\clearpage

\end{document}